\documentclass[%
nofootinbib,
mathtools,
amsmath,
amssymb,
twocolumn,
prl,
longbibliography,
superscriptaddress]{revtex4-2}

\usepackage{booktabs}
\usepackage{mathrsfs}
\usepackage{graphicx}
\usepackage{dcolumn}
\usepackage{bm}
\usepackage{lipsum}
\usepackage{graphics}
\usepackage{bbm}
\usepackage[dvipsnames]{xcolor}
\usepackage{color}
\usepackage{microtype}
\usepackage{IEEEtrantools}
\usepackage{mathrsfs}
\usepackage{amsthm}
\usepackage{enumitem}
\usepackage[normalem]{ulem}
\usepackage[pdftex,breaklinks,colorlinks,
linkcolor=Blue,
citecolor=teal,
anchorcolor=red,
urlcolor=cyan]{hyperref}
\usepackage{orcidlink}
\usepackage{mathtools}
\usepackage{chngcntr}

\newcommand{\e}{\varepsilon}
\newcommand{\lam}{\lambda}
\newcommand{\R}{{\rm R}}
\renewcommand{\O}{\mathcal{O}}

\begin{document}

\title{Black hole mergers beyond general relativity: a self-force approach}

\author{Ayush Roy\,\orcidlink{0009-0006-7597-8566}}
\affiliation{School of Mathematical Sciences and STAG Research Centre, University of Southampton, Southampton, United Kingdom, SO17 1BJ}
\author{Lorenzo K\"uchler\,\orcidlink{0000-0002-6609-1684}}
\affiliation{School of Mathematical Sciences and STAG Research Centre, University of Southampton, Southampton, United Kingdom, SO17 1BJ}
\author{Adam Pound\,\orcidlink{0000-0001-9446-0638}}
\affiliation{School of Mathematical Sciences and STAG Research Centre, University of Southampton, Southampton, United Kingdom, SO17 1BJ}
\author{Rodrigo Panosso Macedo\orcidlink{0000-0003-2942-5080}}
\affiliation{Center of Gravity, Niels Bohr Institute, Blegdamsvej 17, 2100 Copenhagen, Denmark}

\date{\today}

\begin{abstract}
Gravitational waves from binary black hole mergers provide a glimpse of gravitational dynamics in its most extreme observable regime, potentially enabling precision tests of general relativity (GR) and of the Kerr description of black holes. However, until recently, numerical simulations of black hole mergers have not been possible in theories beyond GR. While recent breakthroughs have overcome that obstacle, simulations covering the full, interesting range of binary parameters remain unfeasible. Here we present a new first-principles approach to this problem. We show how self-force theory can be used to model the merger and ringdown of black holes in a broad class of gravitational theories, assuming one object is much smaller than the other. We calculate self-force effects on the merger waveform for the first time, and we demonstrate how our formulation allows us to modularly compute beyond-GR effects and readily incorporate them into a fast merger-ringdown waveform model.
\end{abstract}

\maketitle

\emph{Introduction.}---Over the past decade, general relativity~(GR) has come into its own. For many decades prior, experiments could only directly probe regimes of weak gravity or cosmological scales~\cite{Will:2014kxa,Baker:2014zba}. Now, with the advent of gravitational-wave~(GW) astronomy~\cite{LIGOScientific:2016aoc} and very-long-baseline interferometry~\cite{EventHorizonTelescope:2019dse}, we are able to regularly observe regions of strong gravity through the inspiral and merger of binary black holes~(BHs)~\cite{LIGOScientific:2025slb} and the interaction of BHs with surrounding matter~\cite{EventHorizonTelescope:2022wkp}. These observations have placed new constraints on deviations from GR~\cite{EventHorizonTelescope:2020qrl,Yunes:2013dva,Sanger:2024axs}, and future instruments will provide more stringent tests of GR, including precise measures of whether astrophysical BHs have the unique, Kerr form predicted by GR~\cite{Berti:2015itd,Barausse:2020rsu,Perkins:2020tra,LISA:2022kgy,Vagnozzi:2022moj,Johnson:2024ttr,Abac:2025saz}.

Performing such tests requires accurate theoretical predictions of BH dynamics and their emitted signals both within GR and in theories beyond GR~\cite{Yunes:2009ke,Moore:2021eok, Hu:2022bji, Saini:2023rto, Dhani:2024jja, DuttaRoy:2024aew, Gupta:2024gun, Kapil:2024zdn, Chandramouli:2024vhw, Shen:2025svs}. In the context of GWs emitted by BH binaries, beyond-GR models based on effective-field-theory (EFT) extensions to GR~\cite{Donoghue:1994dn,Donoghue:1995cz,Sotiriou:2011dz,Kovacs:2020pns,Reall:2021ebq,CarrilloGonzalez:2022fwg,Cayuso:2023xbc,Figueras:2024bba,Bernard:2025dyh} have been developed for the binary's inspiral stage, using post-Newtonian theory~\cite{Blanchet:2013haa} in the case of  comparable-mass binaries~\cite{Yagi:2011xp,Sennett:2016klh,Bernard:2018hta,Lang:2013fna, Julie:2019sab, Shiralilou:2021mfl,Bernard:2022noq, Diedrichs:2023foj,Bhattacharyya:2023kbh,Almeida:2024cqz,Trestini:2024mfs} and self-force theory~\cite{Poisson:2011nh,Barack:2018yvs,Pound:2021qin} in the case of binaries with large mass disparities~\cite{Sopuerta:2009iy,Maselli:2020zgv,Maselli:2021men,Barsanti:2022ana,Barsanti:2022vvl,Spiers:2023cva,Speri:2024qak}. Beyond-GR effects in the final ringdown of the post-merger remnant BH have also been studied using BH perturbation theory~\cite{Wagle:2021tam,Hussain:2022ins,Li:2022pcy,DAddario:2023erc,Maenaut:2024oci,Khoo:2024agm,Chung:2024ira,Chung:2024vaf,Chung:2025gyg,Li:2025fci}. However, first-principles study of the merger itself has been restricted to fully nonlinear numerical relativity (NR) simulations of comparable-mass binaries~\cite{Okounkova:2017yby, Silva:2020omi, Witek:2018dmd, Witek:2020uzz, East:2020hgw, Figueras:2021abd, Ripley:2022cdh, Elley:2022ept, Corman:2022xqg,Ma:2023sok, Corman:2024vlk, AresteSalo:2025sxc,Lara:2025kzj}, which have only recently become possible. Such simulations are necessarily limited to assuming particular theories and specifying particular values of physical parameters, with weeks to months of runtime required for each simulation.

In this letter, we describe how the merger and ringdown can be consistently modelled in EFT extensions of GR using self-force theory. Our approach combines the beyond-GR inspiral modelling program of Refs.~\cite{Maselli:2020zgv,Maselli:2021men,Barsanti:2022ana,Barsanti:2022vvl,Spiers:2023cva,Speri:2024qak} with the merger-ringdown modelling program of Ref.~\cite{Kuchler:2025hwx}. Since self-force models are based on expansions in powers of the mass ratio (the mass of the smaller, secondary object over the mass of the larger, primary BH), our method is restricted to binaries with disparate masses. However, self-force models are known to be highly accurate even for currently observable systems with mass ratios $\sim 1:10$ or even closer to unity~\cite{LeTiec:2011bk,LeTiec:2014oez,vandeMeent:2020xgc,Warburton:2021kwk,Wardell:2021fyy,Ramos-Buades:2022lgf,NavarroAlbalat:2022tvh,Albertini:2022rfe}. Moreover, our method represents an entirely new way of exploring the merger regime beyond GR, independent of and disjoint from NR. Our approach also has two key advantages: it allows for exploring entire \emph{classes} of theories simultaneously because higher-curvature terms in an EFT are suppressed by powers of the mass ratio; and it leads to a modular, fast waveform model due to the split between (slow) offline and (fast) online calculations in Ref.~\cite{Kuchler:2025hwx}'s waveform-generation framework.

As a proof of principle, we calculate self-force effects on the merger-ringdown waveform---for the first time in \emph{any} theory---in a broad class of theories involving a nonminimally coupled scalar field.


\emph{Self-force beyond GR.}---From an EFT perspective, GR describes the low-energy limit of some unknown high-energy theory. Beyond leading order in such an EFT, one generically expects higher-order curvature terms and couplings to new fields, correcting the Einstein-Hilbert term in the Lagrangian density: ${\cal L} = {\cal L}_{\rm EH}[{\sf g}] + {\cal L}_{\rm bGR}[\Psi,{\sf g}]$, where ${\sf g}_{\alpha\beta}$ is the spacetime metric and $\Psi$ represents the new fields. References~\cite{Berti:2015itd,Bernard:2025dyh} (for example) survey possible beyond-GR terms ${\cal L}_{\rm bGR}$.

This framework was analysed in the small-mass-ratio limit in Refs.~\cite{Maselli:2020zgv,Spiers:2023cva}. We can organize it in terms of the ratio $\e := \mathring{m}_2/\mathring{m}_1\leq1$ between the two objects' bare masses (the masses of the objects in isolation and in the absence of new fields). A curvature term must come in the form $\propto \alpha^{(q)} {\cal R}^{(q)}$ for some $q$th-order combination of curvature tensors and some set of coupling constants $\alpha^{(q)}\sim \ell^{2q-2}$, where $\ell$ is a length scale that experiments dictate must be comparable to or smaller than a solar mass, $M_\odot$. Hence, given $\mathring{m}_2\gtrsim M_\odot$, the curvature term scales as $\alpha^{(q)} {\cal R}^{(q)}\lesssim \e^{2q-2}/\mathring{m}_1^2$ over most of the spacetime (where ${\cal R}\sim 1/\mathring{m}_1^2$) but as $\alpha^{(q)} {\cal R}^{(q)}\lesssim 1/\mathring{m}_2^2$ very near the secondary (where ${\cal R}\sim 1/\mathring{m}_2^2$), meaning higher-order terms in the EFT are suppressed by higher powers of $\e$ except in a small neighborhood of the secondary. Here we consider the most common type of EFT, involving an additional scalar field~$\varphi$: ${\cal L}_{\rm bGR}=-\frac{\sqrt{{\sf g}}}{8\pi}\bigl[\nabla_{\!\alpha}\varphi\nabla^\alpha\varphi + \alpha^{(2)}F(\varphi){\cal R}^{(2)}+\ldots\bigr]$ (with no scalar potential, for simplicity). This encompasses Einstein-scalar-Gauss-Bonnet, dynamical Chern-Simons (dCS), and a large subset of other Horndeski and quadratic gravity theories~\cite{Berti:2015itd}. 

In the $\e\ll1$ limit, the secondary is also skeletonized---reduced to a point particle with multipole structure---leading to the addition of a term $S_{\rm pp}$ in the action. The new physics in the secondary's vicinity dresses $S_{\rm pp}$ with new parameters such as a scalar charge $q$ as well as corrections to the secondary's bare moments.  On the other hand, the suppression of curvature terms suppresses effects on the primary, endowing it with a negligible scalar charge $Q\sim \alpha^{(2)}\sim \e^2$ and allowing us to approximate its physical mass as $m_1\approx\mathring{m}_1$, for example. Finally, the metric is decomposed as ${\sf g}_{\alpha\beta}=g_{\alpha\beta}+h_{\alpha\beta}$, where $g_{\alpha\beta}$ is the bare metric of the primary BH, which we take to be a Schwarzschild metric of mass $m_1$, and $h_{\alpha\beta}$ is the perturbation arising from the presence of the particle (as well as from  effects of new physics on the primary). In terms of $\e$ and the secondary's charge-to-mass ratio $\lam:=q/\mathring{m}_2$, the fields scale as $h_{\alpha\beta}\sim \e$ and
$\varphi\sim q/m_1 = \lam \e$. 

The field equations, obtained by varying the action with respect to the metric and scalar field~\cite{Spiers:2023cva}, are then\footnote{Strictly speaking, Ref.~\cite{Spiers:2023cva} obtained these field equations from the variational principles $\delta(S_{\rm EH}+S_{\rm bGR})/\delta{\sf g}_{\alpha\beta}+\delta S_{\rm pp}/\delta\tilde g_{\alpha\beta}=0$ and $\delta(S_{\rm EH}+S_{\rm bGR})/\delta\varphi+\delta S_{\rm pp}/\delta\varphi^\R=0$, which will be rigorously derived elsewhere~\cite{Upton:InPrep}. Alternatively, the $\delta$ sources along with the equations of motion~\eqref{eq:xp EOM} and \eqref{eq:m2 EOM} follow from analysis of the field equations in a small region outside the secondary~\cite{Upton:InPrep}, applying the standard method of matched asymptotic expansions~\cite{Mino:1996nk,Gralla:2008fg,Pound:2009sm,Gralla:2010cd,Gralla:2013rwa,Pound:2012dk,Barack:2018yvs,Upton:2021oxf}. See also Refs.~\cite{Galley:2008ih,Zimmerman:2015hua,Cheung:2024byb}. Note finally that our $q$ and $\varphi$ differ by a factor $1/2$ relative to those of Ref.~\cite{Spiers:2023cva} in order to conform to traditional conventions in self-force theory~\cite{Poisson:2011nh}.}
\begin{align}
    \delta G_{\alpha\beta}[h] &= 8 \pi \int m_2\, \tilde u_\alpha \tilde u_\beta\frac{\delta^4(x^\mu - x_p^\mu)}{\sqrt{-\tilde g}}d\tilde\tau - \delta^2 G_{\alpha\beta}[h,h] \nonumber\\ 
    &\quad + 2\Bigl(g_\alpha^{\ \mu}g_\beta^{\ \nu} - \frac{1}{2}g_{\alpha\beta}g^{\mu\nu}\Bigr)\nabla_{\!\mu}\varphi\nabla_{\!\nu}\varphi + \O(3),\!\label{eq:EFE}\\
    \nabla_{\!\alpha}\nabla^\alpha \varphi &= - 4 \pi q \int \frac{\delta^4(x^\mu - x_p^\mu)}{\sqrt{-\tilde g}}d\tilde\tau + \O(2),\label{eq:scalar field EOM}
\end{align}
where we use ``$\O(n)$'' to denote quantities of order $\e^n\lam^k$ (for any $k\geq0$), and we highlight that the coupling constants $\alpha^{(2)}$ only enter in those higher-order terms.\footnote{As a consequence, dCS is indistinguishable from GR at this order because in dCS the secondary will possess a scalar dipole but not a scalar monopole $q$~\cite{Delsate:2018ome,R:2022tqa}.} Here $\delta G_{\alpha\beta}[h]$ is the linearized Einstein tensor on the background $g_{\alpha\beta}$; $\delta^2 G_{\alpha\beta}[h,h]$ is the second-order Einstein tensor, quadratic in $h_{\alpha\beta}$; and the term quadratic in $\varphi$ is the stress-energy tensor of the scalar field. The Dirac $\delta^4$ source terms are the particle's Detweiler stress-energy tensor~\cite{Detweiler:2011tt,Upton:2021oxf} and charge distribution, which we write in terms of the effective metric $\tilde g_{\alpha\beta} = g_{\alpha\beta} + h^\R_{\alpha\beta}$, where $h^\R_{\alpha\beta}$ is a certain smooth vacuum perturbation~\cite{Detweiler:2002mi,Pound:2012dk}.  

At leading order beyond the test-mass approximation, the particle is subject to the standard linear gravitational and scalar self-forces~\cite{Mino:1996nk,Quinn:2000wa,Poisson:2011nh} exerted by $h^\R_{\alpha\beta}$ and the Detweiler-Whiting regular field $\varphi^\R$~\cite{Detweiler:2002mi,Poisson:2011nh,Harte:2014wya}, implying 
\begin{align}
    \frac{D^2x^\alpha_p}{d\tau^2} &=  P^{\alpha\beta}\left[-\frac{1}{2}\bigl(2\nabla_{\!\mu} h^\R_{\nu\beta} - \nabla_{\!\beta} h^\R_{\mu\nu}\bigr)u^\mu u^\nu + \frac{q}{m_2} \nabla_{\!\beta}\varphi^\R\right] \nonumber\\
    &\quad + \O(2),\label{eq:xp EOM}\\
    \frac{dm_2}{d\tau} &= -qu^\alpha\nabla_{\!\alpha}\varphi^\R + \O(3),\label{eq:m2 EOM}
\end{align}
in terms of the proper time $\tau$, four-velocity $u^\alpha:=dx^\alpha/d\tau$, and covariant derivative $\nabla_\alpha$ defined from $g_{\alpha\beta}$, with  $P^{\alpha\beta}\coloneqq g^{\alpha\beta}+u^\alpha u^\beta$. Here the beyond-GR effects arise solely from the presence of the scalar field sourced by the particle; other beyond-GR effects are suppressed, entering at higher orders~\cite{Spiers:2023cva}.

We can further organize these equations by expanding the fields as $h_{\alpha\beta} \hspace{-1pt}=\hspace{-1pt} \e h^{(1,0)}_{\alpha\beta} + \e^2\Bigl(h^{(2,0)}_{\alpha\beta} + \lam^2h^{(2,2)}_{\alpha\beta}\Bigr) + \O(3)$ and $\varphi = \e\lam \varphi_{(1,1)} + \O(2)$. The evolution equations~\eqref{eq:xp EOM} and~\eqref{eq:m2 EOM} then inherit this form. For example, the  right-hand side of Eq.~\eqref{eq:xp EOM} can be written as $f^\alpha = \e\Bigl(f^\alpha_{(1,0)} + \lam^2 f^\alpha_{(1,2)}\Bigr) + \O(2)$, and the solution to Eq.~\eqref{eq:m2 EOM} as $m_2 = \mathring{m}_2\Bigl[1 - \e\lam^2 \varphi^\R_{(1,1)} + \O(2)\Bigr]$. Substituting such expansions into the field equations~\eqref{eq:EFE} and \eqref{eq:scalar field EOM} reduces them to a hierarchy of equations for $h^{(n,k)}_{\alpha\beta}$ and $\varphi_{(n,k)}$. As Ref.~\cite{Kuchler:2025hwx} described in the GR setting, the precise form of these expansions, and the resulting hierarchy of field equations, depends on which regime we consider: the particle's inspiral~\cite{Miller:2020bft}; its transition across the innermost stable circular orbit (ISCO)~\cite{Kuchler:2024esj}; or its final plunge into the primary BH, which produces the merger-ringdown signal. The expansion in each regime is informed by the expansion(s) in neighboring regimes through the method of matched asymptotic expansions.


\emph{Modelling the merger \& ringdown.}---Self-force calculations in the inspiral are based on a multiscale expansion~\cite{Hinderer:2008dm,Miller:2020bft,Wardell:2021fyy,Mathews:2025nyb}, which enables rapid waveform generation~\cite{Katz:2021yft,Pound:2021qin,Chapman-Bird:2025xtd} by splitting the problem in two: a slow, offline stage in which waveform ingredients are pre-computed on the binary's orbital phase space; and a fast, online stage in which the waveform is generated by evolving through the phase space. Reference~\cite{Kuchler:2025hwx} showed how the plunge can be formulated in a post-geodesic (PG) expansion that maintains this offline-online split. Here we explain how that framework extends beyond GR.

\begin{figure}
\includegraphics[width=.375\textwidth]{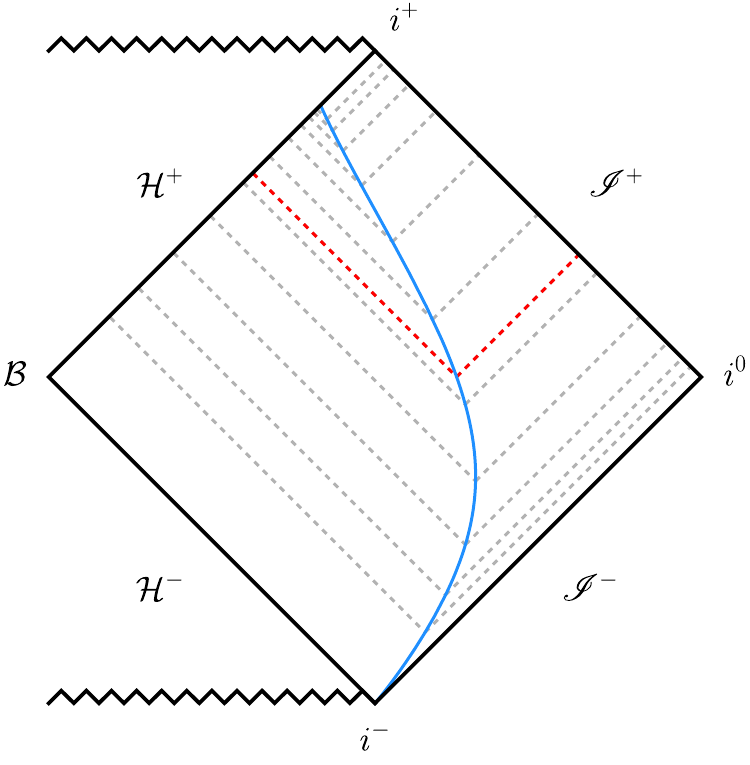}
\caption{Penrose diagram illustrating our choice of spacetime slicing (grey dotted lines) and the particle's geodesic-order plunging trajectory (solid blue curve). The dotted red lines indicate the slice on which the particle crosses the zeroth-order light ring at $r_p=3m_1$. Our choice of slicing links the waveform to the particle at all times along $\mathscr{I}^+$, far after the particle passes the light ring.}
\label{fig:slicing}
\end{figure}

We begin by foliating the spacetime with slices of constant time $s\coloneqq t - \kappa(r_*)$, where $t$ is Schwarzschild time, $\kappa$ is a so-called height function, and $r_*$ is the tortoise coordinate. In our computations, we set $s=t-r_*$ (retarded time $u$) in the region from the particle to $\mathscr{I}^+$, and $s=t+r_*$ (advanced time $v$) in the region from the particle to $\mathcal{H}^+$, as shown in Fig.~\ref{fig:slicing}. At the interface between these regions, we parametrise the particle's worldline as %
\begin{equation}\label{eq:worldline}
    x^\mu_p(t, \e,\lam) = (t, r_p(t, \e,\lam), \theta_p=\pi/2, \phi_p(t, \e,\lam)).
\end{equation}

Here we have restricted to equatorial orbits, without loss of generality. This implies the binary's mechanical phase space has coordinates $(x^A_p, dx^A_p/dt)$,\footnote{The primary BH also acquires dynamical mass and spin corrections $\delta m_1$ and $\delta j_1$ due to absorption of radiation. These contribute $\e^2$ and $\e^2\lambda^2$ terms in the metric, proportional to the fluxes of energy and angular momentum down the horizon. However, they enter as $l=0,1$ terms in the metric, meaning they do not contribute to the waveform (which contains only $l>1$ modes) until they couple to other modes at higher order.} with $A=(r,\phi)$. As in Ref.~\cite{Kuchler:2025hwx}, we further restrict ourselves to plunges that arise from a quasicircular inspiral, in which case the phase space can be reduced to $(\phi_p,d\phi_p/dt)$, with $r_p$ reduced to a function of $\Omega:=d\phi_p/dt$. Also following Ref.~\cite{Kuchler:2025hwx}, we treat $r_p$ as a proxy for $\Omega$, such that $(\phi_p,r_p)$ are the phase-space coordinates. We treat all forces and fields as functions on this phase space and expand them in integer powers of $\e$ while keeping $(\phi_p, r_p)$ fixed. The trajectory through phase space is then governed by a set of simple ordinary differential equations (ODEs):%
\begin{subequations}\label{eq:orbital ODEs}%
\begin{align}
    \frac{d\phi_p}{dt} &= \Omega_{(0)} + \e\bigl[\Omega_{(1,0)} +\lam^2\Omega_{(1,2)}\bigr] + \O(2),
    \\
    \frac{dr_p}{dt} &= F_{(0)} + \e\Bigl[F_{(1,0)}+\lam^2F_{(1,2)}\Bigr] + \O(2),
\end{align}
\end{subequations}
where quantities on the right are functions of $r_p$. The frequencies $\Omega_{(n,k)}$ and forcing functions $F_{(n,k)}$ are obtained from the equation of motion~\eqref{eq:xp EOM} together with early-time conditions arising from the asymptotic match with the transition-to-plunge dynamics. The leading, geodesic (0PG) terms are algebraic~\cite{Kuchler:2025hwx},
\begin{align}\label{eq:0PG dynamics}
    \Omega_{(0)} = \sqrt{\frac{3}{2}}\frac{3m_1}{r_p^2}f_p \quad\ \text{and}\ \quad
    F_{(0)} = -\frac{D_p^{3/2}}{2^{3/2}}f_p,
\end{align}
with $D_p:=(6m_1/r_p - 1)$ and $f_p:=1-2m_1/r_p$, while the first post-geodesic (1PG) terms are governed by ODEs:
\begin{subequations}   \label{eq:1PG dynamics} 
\begin{align} \partial_{r_p}\Omega_{(1,k)}+a_\Omega\,\Omega_{(1,k)} &= b_{\alpha} f^\alpha_{(1,k)},\label{eq:pG1}\\
    \partial_{r_p}F_{(1,k)} +a_{r_p}F_{(1,k)} &= c_{\alpha}f^\alpha_{(1,k)} +d_\Omega\,\Omega_{(1,k)}\label{eq:pG2}.
\end{align} 
\end{subequations}
The coefficients $a_\Omega(r_p)$, $a_{r_p}(r_p)$, $b_\alpha(r_p)$, $c_\alpha(r_p)$, and $d_\Omega(r_p)$ are given in Eq.~(3.48) of Ref.~\cite{Kuchler:2025hwx} and in the Supplementary Material. There, the initial conditions for Eqs.~\eqref{eq:pG1}--\eqref{eq:pG2}, as uniquely determined by the match to the transition-to-plunge dynamics, are also described. 

To express the fields $h_{\alpha\beta}$ and $\varphi$ as functions on phase space, we assume they only depend on time $s$ through a dependence on $(\phi_p,r_p)$. Given that $\phi_p$ is a periodic coordinate, we can then expand the fields in discrete Fourier series as well as in powers of $\e$:
\begin{equation}\label{eq:h expansion}
    h_{\alpha\beta} = \sum_{n=1}^\infty\sum_{k \text{ even}}^{n}\sum_{m\in\mathbb{Z}}\!\e^n\lam^{k} h_{\alpha\beta,m}^{(n,k)}(r_p(s), x^i)e^{-im\phi_p(s)},
\end{equation}
and analogously for the scalar field $\varphi$. 

We extract the GW strain as $h\coloneqq\lim_{r\to\infty}\frac{r}{m_1}h_{\bar m\bar m}$ with $\bar m^\mu=\frac{1}{\sqrt{2}r}(0,0,1,-i\csc\theta)$. After decomposing $h$ into spin-weight $-2$ spherical harmonics, we obtain the waveform modes
\begin{multline}\label{eq:mode amplitudes gravity}
    h_{lm} = e^{-im\phi_p}\Bigl\{\e H^{(1)}_{lm}(r_p) \\  + \e^2\!\left[H^{(2,0)}_{lm}(r_p)+\lam^2H^{(2,2)}_{lm}(r_p)\right] + \O(3)\Bigr\},
\end{multline}
where $H^{(n,k)}_{lm}$ are the spherical-harmonic mode coefficients of $\lim_{r\to\infty}\frac{r}{m_1}h^{(n,k)}_{\bar m\bar m,m}$.

Generating waveforms now comprises (i) pre-computing and storing the frequencies and forcing terms on the right-hand side of Eq.~\eqref{eq:orbital ODEs} and the waveform's mode amplitudes $H^{(1)}_{lm}$ as functions of $r_p$; (ii) solving Eq.~\eqref{eq:orbital ODEs} to evolve through phase space.

In Ref.~\cite{Kuchler:2025hwx} we computed the amplitudes $H^{(1)}_{lm}$ (see also Refs.~\cite{Hadar:2009ip,Hadar:2011vj,Folacci:2018cic,Strusberg:2025qfv}). Here, we compute the beyond-GR 1PG frequency correction $\Omega_{(1,2)}$, forcing function $F_{(1,2)}$,  mass correction, and the most easily accessible contributions to the amplitudes $H^{(2,2)}_{lm}$. Concretely, we compute all 1PG effects that are \emph{linear} in the scalar field, neglecting the more challenging~\cite{Upton:2025bja}, quadratic effects that enter $H^{(2,2)}_{lm}$ through the source term $\propto\nabla_\mu\varphi\nabla_\nu\varphi$ in Eq.~\eqref{eq:EFE}.

\emph{Numerical method.}---The essential ingredients we require are the scalar self-force and correction to the particle's mass. Both of these are obtained from the leading-order solution to Eq.~\eqref{eq:scalar field EOM}. To solve that equation on phase space, we write $\varphi = \e\lambda\varphi_{(1,1)}+\O(2)$ and adopt a convenient variant of the ansatz~\eqref{eq:h expansion}:
\begin{equation}
\label{eq:scalar field ansatz}
    \varphi_{(1,1)} = \sum_{l,m} \sigma R_{lm}(\sigma_p(s),\sigma)Y_{lm}(\theta, \phi)e^{-im \phi_p(s)},
\end{equation}
where $\sigma:=2M/r$ is a compactified radial coordinate and $\sigma_p:=2M/r_p$. We label each time slice, illustrated in Fig.~\ref{fig:slicing}, with the value of $\sigma_p$ at the focal point on the worldline where the lines of constant advanced and retarded time meet. $\sigma_p$, which runs from $1/3$ to $1$ during the plunge, then serves as a compactified time coordinate. 

When substituting the ansatz~\eqref{eq:scalar field ansatz} into the field equation~\eqref{eq:scalar field EOM}, we apply the chain rules $\frac{d\sigma_p}{ds} = -\frac{\sigma^2_p}{2M}\frac{dr_p}{dt}\frac{dt_p}{ds}$ and $\frac{d\phi_p}{ds} = \frac{d\phi_p}{dt}\frac{dt_p}{ds}$ along with the leading-order expressions $\frac{dr_p}{dt}=F_{(0)}$, $\frac{d\phi_p}{dt}=\Omega_{(0)}$, and 
\begin{equation}
\frac{dt_p}{ds} = \frac{1}{1 + 2^{-3/2}(3 \sigma_p-1)^{3/2}H}  
\end{equation}
with $H:=d\kappa/dr_*$ equal to $+1$ ($s=u$) or $-1$ ($s=v$).

Following these substitutions, the exponential $e^{-im\phi_p}$ factors out of the field equation as it also appears in the source term in Eq.~\eqref{eq:scalar field EOM}, each $lm$ mode of which is proportional to $\delta(\sigma-\sigma_p)e^{-im\phi_p}$. At points $\sigma \neq \sigma_p$ the numerical variable $R_{lm}$ in the expansion \eqref{eq:scalar field ansatz} then satisfies the homogeneous equation\footnote{We note that due to our choice of characteristic slicing, this equation is first order in time and requires only characteristic, rather than Cauchy, initial data at $\sigma_p=1/3$.} 
\begin{equation}
\label{eq:homogeneous equation field}
    B \partial^2_\sigma R_{lm} + C \partial_{\sigma_p, \sigma} R_{lm} + E \partial_\sigma R_{lm} + VR_{lm} = 0, 
\end{equation}
with coefficients $B = (1 - \sigma) \sigma^2$, $C = 4MHd\sigma_p/ds$, $E = [(2 - 3 \sigma)\sigma - 4imMHd\phi_p/d s]$, and $V = - [l(l+1)+\sigma]$, where we use the leading-order expressions for $d\sigma_p/ds$ and $d\phi_p/ds$ above. Jump conditions at $\sigma=\sigma_p$ are dictated by the $\delta(\sigma-\sigma_p)$ source together with the transformation between slicing from $s=u$ to $s=v$ across the particle.

We solve the homogeneous equation~\eqref{eq:homogeneous equation field} on either side of the particle, together with the jump conditions, by employing a spectral method similar to Ref.~\cite{Canizares:2009ay,Canizares:2010yx}'s (see also \cite{Field:2009kk,Zenginoglu:2011zz,Harms:2014dqa,PanossoMacedo:2014dnr,Thornburg:2016msc,OBoyle:2022yhp,OBoyle:2023jqo,Vishal:2023fye,DaSilva:2023xif,DaSilva:2024yea,Vishal:2025pqc} for related time-domain methods). We divide the spatial domain into $N$ cells on either side of the particle, with the particle always at a cell interface, and expand $R_{lm}$ in each cell as 
\begin{equation}
\label{eq: spectral expansion}
R_{lm}(\sigma_p, \sigma) = \sum_{i=0}^{j-1} c^{lm}_i(\sigma_p)T_i\bigl(\chi(\sigma_p, x(\sigma_p, \sigma))\bigr). 
\end{equation}
Here $T_i$ are Chebyshev polynomials of the first kind, the coefficients $\{c^{lm}_i\}$ are unknown functions of the compact time $\sigma_p$, $x$ linearly maps the range of $\sigma$ in each cell to the range $-1$ to $+1$, and $\chi\in[-1,1]$ is a refinement of $x$ described below. For example, if there is only one cell on either side of the particle, $x(\sigma_p, \sigma)$ takes the form $x = \frac{2(\sigma - \sigma_p)}{\sigma_p} + 1$ for the cell ranging from $\mathscr{I}^+$ to the particle, and $x = \frac{2(\sigma - 1)}{1-\sigma_p}+1$ for the cell ranging from the particle to $\mathcal{H}^+$. Our code can flexibly accommodate different numbers of cells of varying sizes; in our computations for this Letter, we use $j=50$ and $N=5$ with narrower cells closer to the particle. The coordinate $\chi$ is introduced because, as the particle approaches the horizon $(\sigma_p \rightarrow 1^-)$, the numerical variable develops a steep gradient in the limit $\sigma \rightarrow \sigma_p^-$. We resolve this feature using an analytical mesh refinement in the cells adjacent to the particle, explicitly given by Eq.~(96) from Ref.~\cite{PanossoMacedo:2022fdi}.

By evaluating the system of homogeneous equations and jump conditions on Gauss-Lobatto nodes, we obtain a linear algebro-differential problem of the form $\bm{L} \dot{\bm{c}} = \bm{R}\bm{c}$, where $\bm{L}$ and $\bm{R}$ are square matrices of size $2Nj$, and the dot denotes differentiation with respect to $\sigma_p$ at fixed $\chi$. A linear system such as this can be solved by an implicit solver without resorting to a root-finding algorithm, thereby evading the Courant-Friedrichs-Lewy stability condition while maintaining the speed of an explicit solver. In this work, we use the Singly Diagonally Implicit Runge-Kutta scheme of order 3~\cite{Alexander:1977}. Lastly, we choose a time step $\Delta \sigma_p$ that starts at $10^{-3}$ at the ISCO and shrinks to zero as $\sigma_p$ approaches 1: $\Delta \sigma_p=\frac{3}{2}(1-\sigma_p)\times 10^{-3}$.

We use this spectral machinery to solve for $R_{lm}$ up to $l=50$ and $|m|\leq l$. From the field modes $R_{lm}$, we compute the regular field $\varphi^\R_{(1,1)}$ and scalar self-force $f^\alpha_{(1,2)}$ as functions of $\sigma_p$ along the particle's trajectory, using the standard mode-sum method~\cite{Barack:2009ux,Heffernan:2012su}. We find that 50 $l$-modes suffice to calculate these quantities to at least two digits of accuracy up until $\sigma_p \approx 0.99$. Due to strong beaming of the scalar field as the particle approaches the horizon, higher $l$ modes would be required after that time. Errors from our spectral method are far below the error from the $\ell$-mode truncation, with maximum relative error of $\sim 10^{-6}$ in $\varphi^\R_{(1,1)}$ and $f^\alpha_{(1,2)}$ due to truncation of the spectral index at $j=50$.


\begin{figure}[tb]
\includegraphics[width=\columnwidth,trim={0 10pt 0 0}]{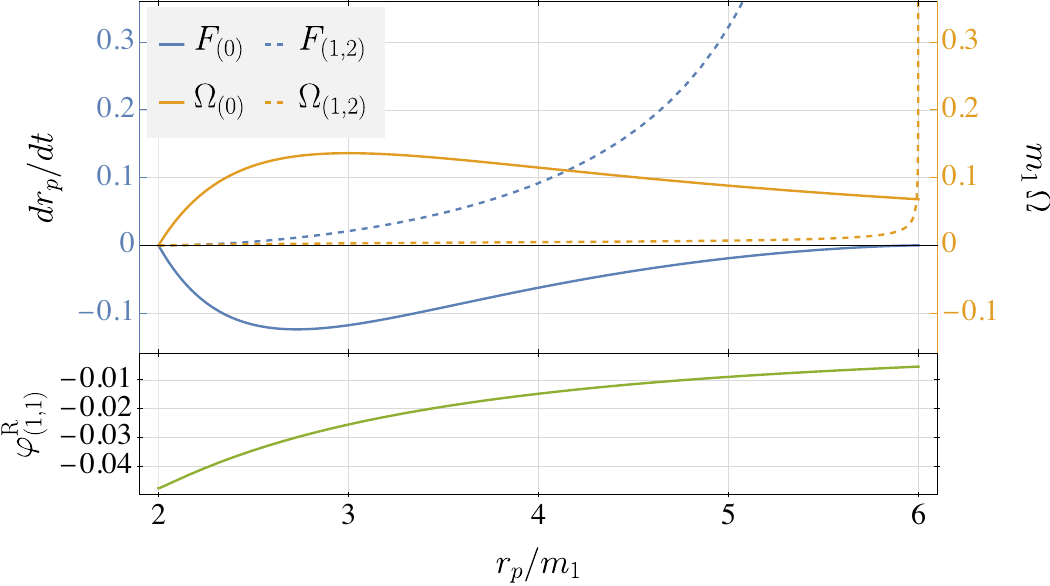}
\caption{Top panel: orbital frequency and rate of change of the orbital radius as a function of the orbital radius itself as the particle plunges from the ISCO at $r_p=6\,m_1$ to the horizon at $r_p=2\,m_1$. Solid curves correspond to 0PG terms in $\Omega$ and $dr_p/dt$ (respectively, $\Omega_{(0)}$ and $F_{(0)}$); dashed curves, to 1PG corrections ($\Omega_{(1,2)}$ and $F_{(1,2)}$). Bottom panel: the regular field $\varphi^\R_{(1,1)}$ as a function of $r_p$.}
\label{fig:rpdot and Omega vs rp}
\end{figure}

\emph{Results.}---Using our data for the self-force $f^\alpha_{(1,2)}$, we first solve the ODEs~\eqref{eq:pG1} and~\eqref{eq:pG2} for the frequency correction $\Omega_{(1,2)}$ and forcing function $F_{(1,2)}$, with initial conditions (asymptotically near the ISCO) determined  from the transition-to-plunge solution. Figure~\ref{fig:rpdot and Omega vs rp} displays the results, with the geodesic frequency and forcing function also shown for comparison. Both 0PG and 1PG quantities go to zero at the horizon due to the infinite redshift there; conversely, the 1PG terms blow up at the ISCO due to the breakdown of the PG approximation there (where the transition-to-plunge dynamics takes over~\cite{Kuchler:2025hwx}). In the lower panel of Fig.~\ref{fig:rpdot and Omega vs rp}, we plot $\varphi^\R_{(1,1)}$, which directly governs the correction $-\mathring{m}_2\e\lam^2 \varphi^\R_{(1,1)}$ to the particle's mass. The plot indicates growth of the mass during the plunge, consistent with Ref.~\cite{Wittek:2024gxn}. Our data for $\varphi^\R_{(1,1)}$ and $f^\alpha_{(1,2)}$ is openly available in Ref.~\cite{data}.

At geodesic order, the orbital frequency follows a familiar pattern~\cite{Damour:2007xr}, climbing from $m_1\Omega_{(0)}=1/(6\sqrt{6})$ at the ISCO to a peak value $1/(3\sqrt{6})$ at the light ring ($r_p=3m_1$) before falling to zero. From our data for $\Omega_{(1,2)}$, we find that the 1PG peak orbital frequency is shifted to $m_1\Omega_{\rm peak} \approx 1/(3\sqrt{6})+0.0029(8)\e\lam^2$.

\begin{figure}[tb]
\includegraphics[width=\columnwidth,trim={0 15pt 45pt 0}]{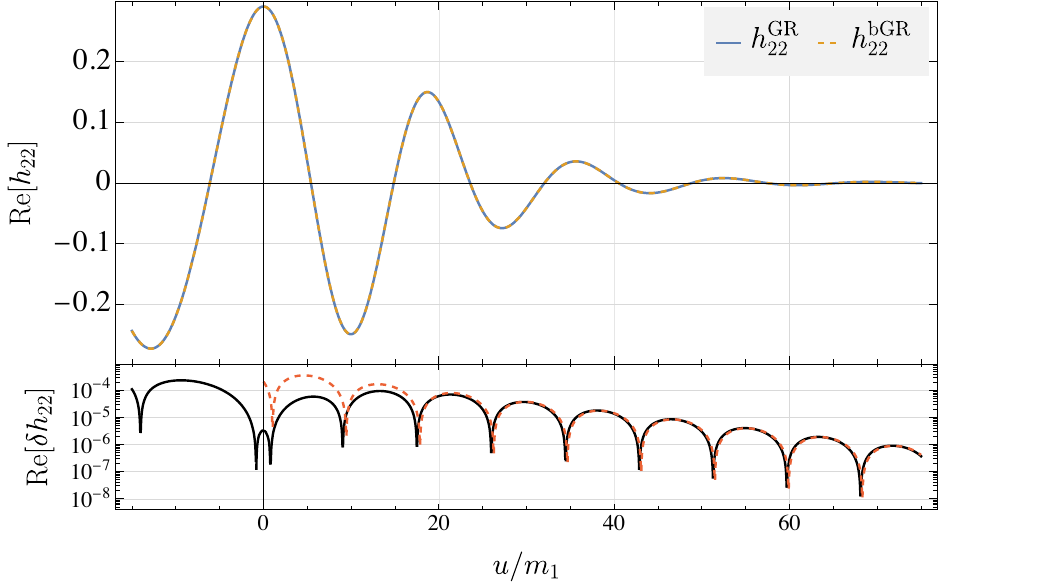}
\caption{Top panel: $(2,2)$ mode of the merger-ringdown waveform in GR and beyond GR, with mass ratio $\varepsilon=1/5$ and charge-to-mass ratio $\lambda=1/8$. The two waveforms, which are aligned in time and phase at the peak of $|h_{22}|$, are indistinguishable on this scale. Bottom panel: the difference $\delta h_{22}$ between the GR and beyond-GR waveforms (black curve) and the $\varepsilon^2\lambda^2$ term in the fundamental QNM (red dashed curve).}
\label{fig:waveform}
\end{figure}

Given the corrections to the particle's mass and trajectory, we can compute their contributions to the waveform amplitudes $H^{(2,2)}_{lm}$ using the same method as for $H^{(1)}_{lm}$~\cite{Kuchler:2024esj}. The precomputed data for $\Omega_{(1,2)}$, $F_{(1,2)}$, and $H^{(2,2)}_{lm}$ then allow us to immediately generate waveforms for any values of $\varepsilon$ and $\lambda$ by solving Eqs.~\eqref{eq:orbital ODEs} and evaluating Eq.~\eqref{eq:mode amplitudes gravity} (neglecting the unknown 1PG GR terms $\Omega_{(1,0)}$, $F_{(1,0)}$, and $H^{(2,0)}_{lm}$). We display an example in Fig.~\ref{fig:waveform}. Our complete set of offline calculations required roughly 40~CPU~hours, after which we can generate a waveform for any $\varepsilon$ and $\lambda$ in roughly 1s in Mathematica on a laptop; we expect this to be reduced to tens of milliseconds in a GPU implementation in the \texttt{FastEMRIWaveforms} package~\cite{Katz:2021yft,Chapman-Bird:2025xtd,FEW}, based on the timing of complete inspiral-merger-ringdown waveforms (built from Refs.~\cite{Wardell:2021fyy,Kuchler:2024esj,Kuchler:2025hwx}) that have been implemented in that package~\cite{Chapman-Bird:private}.

The corrections to the orbital evolution and particle mass also lead to corrections to the waveform's peak amplitude as well as to the amplitudes of the quasinormal modes (QNMs) in the ringdown. By generating waveforms for a sequence of $\lambda$ values, measuring the peak amplitude for each, and fitting the result to an even polynomial in $\lambda$, we find the peak of the (2,2) mode is $|h_{22}|_{\rm max} \approx \varepsilon[1.45 + 5.0(8)\times 10^{-3}\varepsilon\lambda^2]$. We extract the correction to the amplitude of the fundamental QNM in the same way, after fitting the ringdown of each waveform with the form $h_{lm}=\sum_{n}A_{lmn}e^{-i\hspace{.8pt}\omega_{lmn}(u-u_{\rm peak})}$ (with complex $\omega_{lmn}$). We find the amplitude of the fundamental mode is corrected to $A_{220}\approx\varepsilon\{2.36+3.65i + [0.33(0)-0.80(7)i]\varepsilon\lambda^2\}$; the leading, GR term is in agreement with the first-principles (not fitted) value from Ref.~\cite{Kuchler:2025hwx}. Corrections to the QNM frequencies themselves, which would be most relevant to the BH spectroscopy program~\cite{Berti:2025hly}, would arise from the quadratic scalar source in Eq.~\eqref{eq:EFE}, which we have not computed. However, corrections to the amplitudes could be relevant to inspiral-merger-ringdown consistency tests of GR~\cite{Ghosh:2017gfp, Carson:2019kkh}.


\emph{Discussion.}---Our framework provides a new avenue to model BH mergers beyond GR, allowing modular calculations of beyond-GR effects and rapid generation of merger-ringdown waveforms. Using our formalism, we have computed scalar-field corrections to the merger-ringdown waveform for quasicircular, nonspinning BH binaries in a wide class of EFT extensions to GR. We have also computed an invariant dynamical effect (the correction to the peak orbital frequency) and corrections to the waveform's peak and ringdown. 

As we stressed above, our results do not represent a complete accounting of the leading deviations from GR, as we have not computed one of the $\e^2\lam^2$ terms, whose relative magnitude cannot be predicted in advance. However, our work represents a proof of principle that such calculations are feasible. (See also Ref.~\cite{OuldElHadj:2024psw}.) Our demonstration applies to most theories of interest involving an additional scalar field, and it can be readily extended to theories with additional vector fields or higher-order curvature terms. The main restriction in our approach is that the primary black hole cannot differ significantly from a Kerr black hole (i.e., by corrections larger than the mass ratio, as can occur for a scalar-tensor theory whose potential term violates energy conditions, for example~\cite{Herdeiro:2015waa}). 

When combined with a model of the inspiral~\cite{Maselli:2020zgv,Miller:2020bft,Wardell:2021fyy,Spiers:2023cva} and transition to plunge~\cite{Kuchler:2024esj}, our formalism can be employed to build beyond-GR inspiral-merger-ringdown waveform models in self-force theory. Alternatively, it can inform parametrized and theory-specific effective-one-body models~\cite{Silva:2022srr, Pompili:2025cdc, Julie:2024fwy}, as has been done with self-force results for the inspiral within GR~\cite{Nagar:2022fep,vandeMeent:2023ols, Leather:2025nhu}.

We also emphasize that, although we have focused on their beyond-GR aspects, our results represent the first-ever calculations of self-force effects in BH merger-ringdown waveforms. Hence, they also serve as the first step toward the development of 1PG waveforms \emph{within} GR. Followup work will take the next step by computing the first-order \emph{gravitational} self-force in the plunge.

\emph{Acknowledgments.}---AP thanks Suvendu Giri, Luis Lehner, Andrea Maselli, Mostafizur Rahman, Thomas Sotiriou, Nico Yunes, and especially Andrew Spiers for helpful discussions. AR and AP acknowledge the support of a Royal Society University Research Fellowship. LK and AP acknowledge the support of the ERC Consolidator/UKRI Frontier Research Grant GWModels (selected by the ERC and funded by UKRI [grant number EP/Y008251/1]). %
R.P.M. acknowledges support from the Villum Investigator program supported by the VILLUM Foundation (grant no. VIL37766) and the DNRF Chair program (grant no. DNRF162) by the Danish National Research Foundation. The Center of Gravity is a Center of Excellence funded by the Danish National Research Foundation under grant No. 184.
This work makes use of the Black Hole Perturbation Toolkit.

\bibliography{bibfile}

\clearpage


\section*{Supplementary Material}
\setcounter{page}{1}
\setcounter{equation}{0}
\setcounter{figure}{0}
\renewcommand{\thefigure}{S\arabic{figure}}
\renewcommand{\theequation}{S\arabic{equation}}

\subsection*{Asymptotic boundary conditions for the plunge dynamics}

In this supplementary material we explain how the 1PG plunge dynamics are uniquely determined by Eq.~\eqref{eq:1PG dynamics} of the Letter once asymptotic boundary conditions are provided from the transition to plunge. The content here largely duplicates Ref.~\cite{Kuchler:2025hwx} but is included for convenience.

The 1PG corrections to the angular velocity $d\phi_p/dt=\Omega$ and radial velocity $dr_p/dt=F^{r_p}$ are governed by Eq.~(3.48) of Ref.~\cite{Kuchler:2025hwx}, reproduced as Eq.~\eqref{eq:1PG dynamics} in the main text of the Letter. Explicitly, 
\begin{widetext}
\begin{subequations}\label{eq:1PG dynamics supplementary}
\begin{align}\label{eq:1PG Omega}
    \frac{\partial\Omega_{(1,k)}}{\partial {r_p}}+\frac{2(r_p-3M)}{r_p(r_p-2M)}\Omega_{(1,k)} &= -\frac{\sqrt{6}}{8M}f^r_{(1,k)}-\frac{\sqrt{3}(r_p-2M)(2r_p^3-27M^2 r_p+54M^3)}{4Mr_p^{5/2}(6M-r_p)^{3/2}} f^t_{(1,k)},\\
    \frac{\partial F_{(1,k)}}{\partial{r_p}} - \frac{M(5r_p+6M)}{r_p(r_p-2M)(6M-r_p)}F_{(1,k)} &= -\frac{12\sqrt{3}M(r_p-2M)}{r_p^{1/2}(6M-r_p)^{3/2}}\Omega_{(1,k)} -\frac{9(r_p-2M)^2}{8r_p^2}f^t_{(1,k)}\nonumber\\
    &\quad -\frac{9\sqrt{2}r_p^{1/2}(r_p-2M)}{4(6M-r_p)^{3/2}}f^r_{(1,k)},\label{eq:1PG Fr}
\end{align}
\end{subequations}
\end{widetext}
for $k=0,2$. Here 
\begin{equation}    
f^\alpha_{(1,0)}=-\frac{1}{2}P^{\alpha\beta}_{(0)}\bigl(2\nabla_{\!\mu} h^{\R(1,0)}_{\nu\beta} - \nabla_{\!\beta} h^{\R(1,0)}_{\mu\nu}\bigr)u^\mu_{(0)} u^\nu_{(0)} 
\end{equation}
and 
\begin{equation}
f^\alpha_{(1,2)}=P^{\alpha\beta}_{(0)}\nabla_\beta\varphi^{\rm R}_{(1,1)} 
\end{equation}
are the coefficients of $\varepsilon\lambda^0$ and $\varepsilon\lambda^2$ in the gravitational and scalar self-accelerations appearing in Eq.~\eqref{eq:xp EOM} of the Letter. Here $u^\alpha_{(0)}$ and $P^{\alpha\beta}_{(0)}=(g^{\alpha\beta}+u^\alpha_{(0)} u^\beta_{(0)})$ are evaluated as functions of $r_p$ using the 0PG dynamics. Equation~\eqref{eq:1PG dynamics supplementary} was derived in Ref.~\cite{Kuchler:2025hwx} in the case of the gravitational self-force, but the derivation is identical for the scalar self-force.

Equation~\eqref{eq:1PG dynamics supplementary} requires initial conditions asymptotically near the ISCO, $r_p=6M$. These are determined from the requirement that the 1PG dynamics asymptotically matches the transition-to-plunge dynamics, which governs the passage across the ISCO. In the transition to plunge, in a small range of $r_p$ around the ISCO, we work with a scaled variable $\Delta r_p\coloneqq (r_p-6M)/\varepsilon^{2/5}$, which is of order unity in the transition regime $|r_p-6M|={\cal O}(\varepsilon^{2/5})$~\cite{Buonanno:2000ef,Ori:2000zn,Kuchler:2024esj}. If the PG dynamics are expanded for small $|r_p-6M|$ and the transition dynamics are re-expanded for small $\varepsilon$ at fixed $|r_p-6M|$, in both cases we obtain double series for small $\varepsilon$ and small $|r_p-6M|$. These two double series must match order by order in each of the two small quantities. As explained in Ref.~\cite{Kuchler:2025hwx}, this criterion provides asymptotic boundary conditions for $\Omega_{(0)}$, $F_{(0)}$, $\Omega_{(n,k)}$, and $F_{(n,k)}$ in the limit $r_p\to 6M$. 

This matching condition fully fixes the 0PG dynamics to take the algebraic form~\eqref{eq:0PG dynamics}, and it provides boundary conditions for Eq.~\eqref{eq:1PG dynamics supplementary}. The latter are given by Eqs.~(3.50) and~(3.51) of Ref.~\cite{Kuchler:2025hwx}, reproduced here in the notation of the Letter:
\begin{widetext}    
\begin{subequations}\label{eq:1PG ISCO condition}
\begin{align}
     \Omega_{(1,k)} &= -\frac{9\sqrt{2}M^{1/2}f^{t\star}_{\!(1,k)}}{(6M-r_p)^{1/2}} - \frac{3}{4}\sqrt{\frac{3}{2}}f^{r\star}_{(1,k)} - \frac{3 \left(11f^{t\star}_{(1,k)} + 12 M \partial_{r_p}f^{t\star}_{(1,k)} \right)}{2 \sqrt{2} M}(6M-r_p)^{1/2} + {\cal O}(6M-r_p),\\
\label{eq:Fr1 asymp}
    F_{(1,k)} &= -\frac{864M^2 f^{t\star}_{(1,k)}}{(6M-r_p)} + 24M \left(f^{t\star}_{(1,k)} - 12M \partial_{r_p}f^{t\star}_{(1,k)}\right) + {\cal O}[(6M-r_p)^{1/2}],
\end{align}
\end{subequations}
\end{widetext}
where we use a star to denote evaluation at $r_p=6M$. Notice that these expressions diverge at the ISCO, indicating the breakdown of the PG expansion there; as indicated in the body of the Letter, that breakdown is cured by instead using the transition-to-plunge dynamics in a vicinity of the ISCO. However, we stress that the plunge dynamics is fully determined by Eqs.~\eqref{eq:1PG dynamics supplementary} and \eqref{eq:1PG ISCO condition}, and the resulting merger-ringdown waveform is independent of how one splices the transition-to-plunge and plunge in a complete inspiral-merger-ringdown model. Equation~\eqref{eq:1PG ISCO condition} represents the sole input from the transition to plunge.

We solve Eq.~\eqref{eq:1PG dynamics supplementary} subject to Eq.~\eqref{eq:1PG ISCO condition} in the specific case $k=2$ (i.e., with the scalar rather than gravitational self-force). To do so, we work with the rescaled variables $
\overline{\Omega}_{(1,2)} \coloneqq (6M-r_p)^{1/2}\Omega_{(1,2)}$ and $\overline{F}_{(1,2)} \coloneqq (6M-r_p)F_{(1,2)}$, which tend to constants at the ISCO. Even so, the ODE system remains singular at the points $r_p=6M,2M$, with factors of $(6M-r_p)$ and $(r_p-2M)$ multiplying derivatives (after multiplying through by the highest powers of these factors to remove them from  denominators). This poses numerical challenges for simple ODE solvers, even using implicit or spectral techniques. However, we found the ODEs can be solved using Mathematica's built-in NDSolve function with higher than machine precision over the restricted domain $r_p=\{2+10^{-x}, 6-10^{-x}\}$ with $x=\{4,6,8,10,12\}$ and with boundary conditions at $r_p=6-10^{-x}$ provided by Eq.~\eqref{eq:1PG ISCO condition}. Except very near the ISCO, the resulting solution exhibited at least 3 digits of agreement irrespective of the value of $x$ (and with uncertainties subdominant to those from our truncation of the $l$-mode sum when calculating the self-force). Results in the Letter use data with $x=12$.

\subsection*{Validation/convergence tests}

In this section we validate our numerical implementation and illustrate the convergence tests on which most of our error estimates are based. We test convergence with respect to the time step, convergence of the spectral (Chebyshev) expansion, and convergence of the sums over~$l$.

Figure~\ref{fig:1,1_particleloc_timestep_convergence} shows the convergence with step size of our time evolution. In our simulations we use a time step adapted to the slow changes of our numerical field at early times and rapid change at late times: 
\begin{equation}\label{dsigmap}
\Delta \sigma_p = \frac{3}{2}(1-\sigma_p)\times 10^{-3}.
\end{equation}
This represents our highest resolution. In Fig.~\ref{fig:1,1_particleloc_timestep_convergence} we multiply this step size by an overall constant and measure the resulting error in our numerical variable relative to our highest resolution. While the relative error is seen to grow over the evolution, it remains small and decreases with decreasing step size, as expected. We also emphasize that our time variable is compact, meaning the error does not grow beyond the amount shown in the plot.

\begin{figure}[tb!]
    \centering
    \includegraphics[width=\columnwidth,trim={0  8pt 0 0},clip]{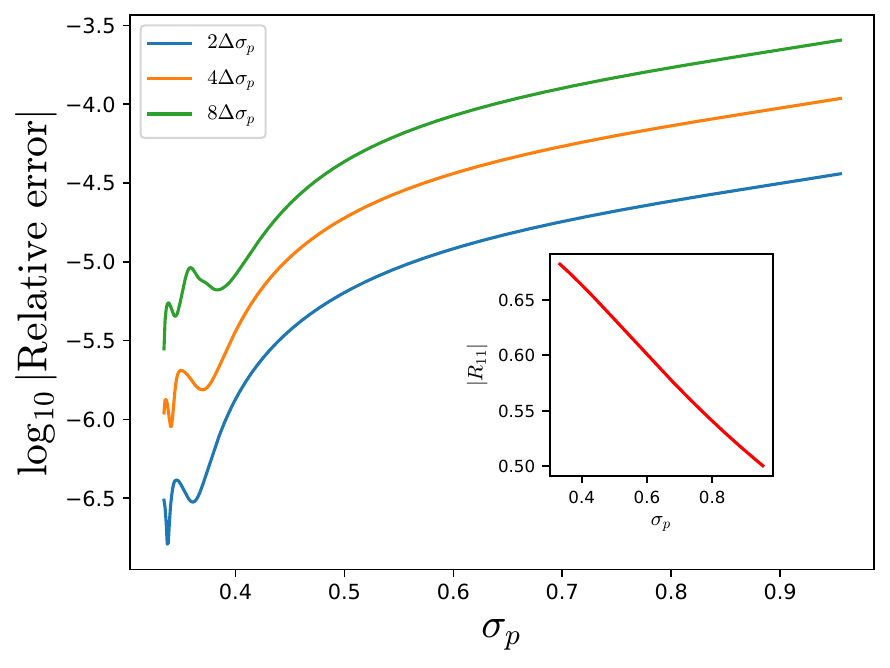}
    \caption{Relative error in the $(l,m)=(1,1)$ component of the field $R_{lm}$ [defined in Eq.~\eqref{eq:scalar field ansatz}] as a function of time at the particle`s location. Each curve represents the relative error for a time step that is a fixed constant multiple of the time step~\eqref{dsigmap} used in the body of the Letter, as displayed in the legend. We observe that the errors decrease as the fixed constant is decreased. The inset shows the absolute value of $R_{11}$ at the particle as a function of compact time.}
    \label{fig:1,1_particleloc_timestep_convergence}
\end{figure}

\begin{figure}[tb!]
    \centering
    \includegraphics[width=\columnwidth,trim={0  10pt 0 0},clip]{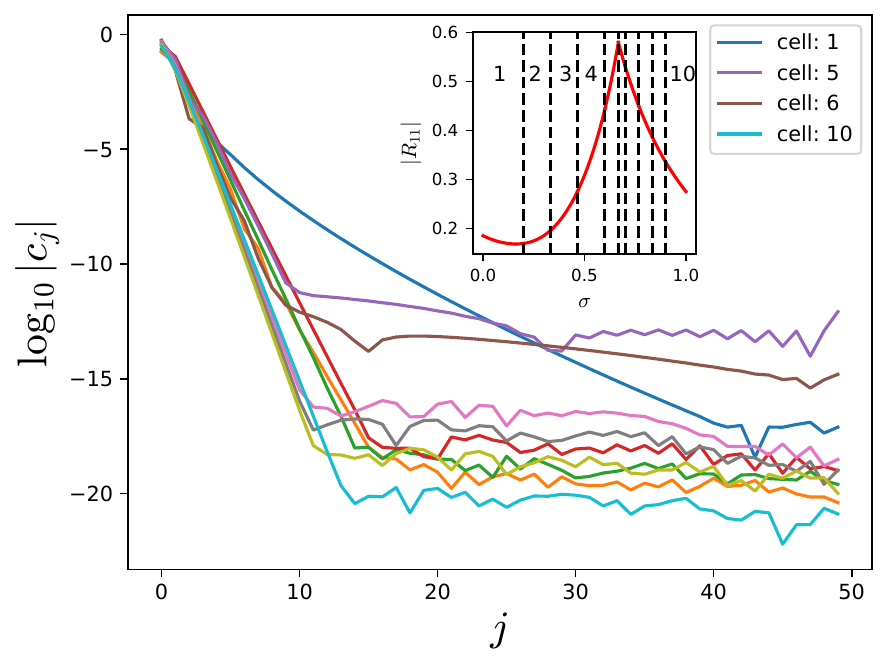}
    \caption{Spectral coefficients in the solution for the $(l,m)=(1,1)$ field, as defined in Eq.~\eqref{eq: spectral expansion}, at time $\sigma_p=1/3$, when the particle is at the light ring. Each curve represents the coefficients in one of the 10 cells into which we subdivide our spatial domain; the division into 10 cells is displayed in the inset, where we plot the absolute value of $R_{11}$ over the full spatial domain. The dashed black lines demarcate cells, five of which are numbered for reference, with numbering increasing from 1 at $\mathscr{I^+}$ to 10 at the horizon.}
    \label{fig:1,1_lightring_N_convergence}
\end{figure}

\begin{figure}[tb!]
    \centering
    \includegraphics[width=\columnwidth,trim={0  8pt 0 0},clip]{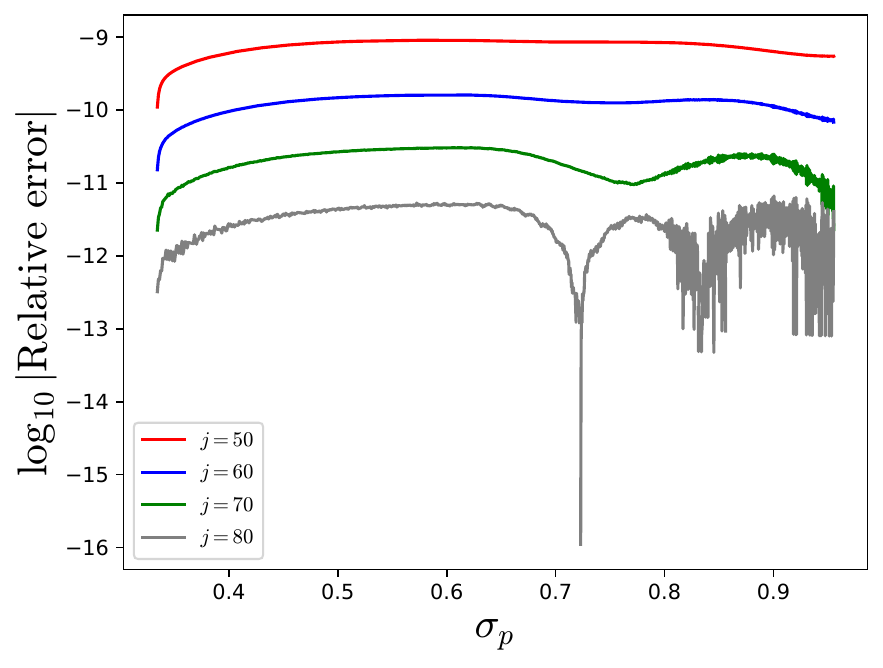}
    \caption{Relative error in the regular field $\varphi^{\rm R}_{(1,1)}$ on the particle as a function of time $\sigma_p$ for various spectral resolutions. Each curve corresponds to a fixed number $j$ of polynomials in each cell, and the error is measured relative to our highest resolution, $j=90$.}
    \label{fig:regularfield_N_convergence}
\end{figure}

\begin{figure}[tb!]
    \centering
    \includegraphics[width=\columnwidth,trim={0  10pt 0 -10pt},clip]{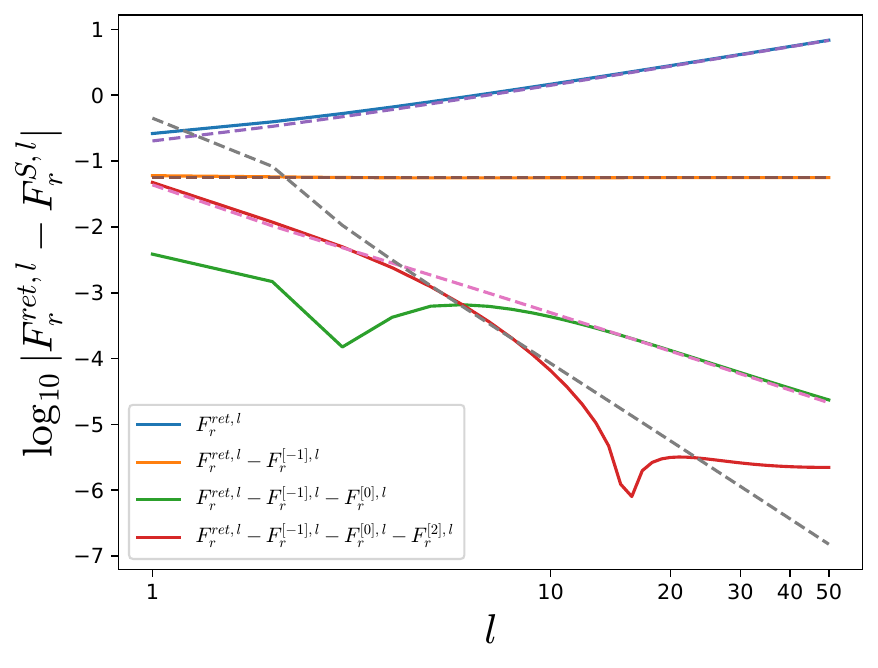}
    \caption{Convergence of the radial self-force with spherical-harmonic mode $l$ when the particle is at the light ring. Each solid line corresponds to a cumulative degree of regularization in the mode sum~\eqref{Fr mode sum}, whereas each dashed line corresponds to individual regularization parameters $F^{[n,l]}_r$. After each successive subtraction, the solid line is expected to agree at large $l$ with the dashed line corresponding to the next regularization parameter. For example, the leading-order-in-$l$ behaviour of the thick green line is captured by the dashed pink line.}
    \label{fig:radialselfforce_lightring_ell_convergence}
\end{figure}

Next, in Figs.~\ref{fig:1,1_lightring_N_convergence} and~\ref{fig:regularfield_N_convergence} we show the convergence of our spectral method. Figure~\ref{fig:1,1_lightring_N_convergence} displays the absolute value of the spectral coefficients specifically for the $(l,m)=(1,1)$ mode of the field $R_{lm}$ when the particle is at the light ring. The 10 curves correspond to the coefficients in the 10 cells used in this work, with cell 1  adjacent to $\mathscr{I^+}$ and cell 10  adjacent to the horizon. Exponential fall-off is observed in all 10 cells, though with a slower rate in cell~1 (consistent with the behavior in frequency-domain results for circular orbits in Fig.~7 of Ref.~\cite{PanossoMacedo:2022fdi}). The coefficients in cell~5 and~6, which are adjacent to the particle, plateau at a relatively high error floor due to strong gradients in the field there at mid to late times, as mentioned in the main text. 

Next, Fig.~\ref{fig:regularfield_N_convergence} shows the convergence of the regular field $\varphi^{\rm R}_{(1,1)}$ with increasing spectral resolution. Here we observe that the error decreases exponentially as $j$ is increased.

Finally, we check the convergence of our sum over $l$ modes. We compute the regular field and self-force using the standard method of mode-sum regularization~\cite{Barack:2009ux}, in which the Detweiler-Whiting singular field is removed from the retarded field $l$ mode by $l$ mode before summing over modes. For example, in the case of the radial self-force, we write~\cite{Heffernan:2012su}
\begin{align}
    F_r &= \sum_l F^l_r \label{Fr mode sum}\\
    &= \sum_l \left(F^{{\rm ret}, l}_r - F_r^{[-1,l]} - F_r^{[0,l]} - F_r^{[2,l]} - F_r^{[4,l]}\right).\nonumber
\end{align}
Here $F^{{\rm ret}, l}_r$ is a radial derivative of the retarded scalar field $l$ mode. Concretely, $\varphi_{lm}=\sigma R_{l m} Y_{lm}e^{-im\phi_p}$ is the individual $(l,m)$ summand in Eq.~\eqref{eq:scalar field ansatz}, and
\begin{equation}
F^{ret,l}_r(r_p) = \partial_r \varphi_{l}(r_p) = \sum_{m=-l}^{+l}\partial_r \varphi_{lm}\Bigr|_{\substack{r=r_p\\\theta=\pi/2\\\phi=\phi_p}}\;.
\end{equation}
The subtracted quantities $F_r^{[n,l]}$, referred to as regularization parameters, are obtained analytically in Ref.~\cite{Heffernan:2012su} from the mode decomposition of the singular field. They diverge at large $l$ or decay with characteristic power laws:%
\begingroup\allowdisplaybreaks
\begin{align}
    F^{[-1,l]}_r &= (2l+1)F_r^{[-1]},\\
    F^{[0,l]}_r &= F^{[0]}_r,\\
    F^{[2,l]}_r &= \frac{F^{[2]}_r}{(2l-1)(2l+3)},\\
    F^{[4,l]}_r &= \frac{F^{[2]}_r}{(2l-3)(2l-1)(2l+3)(2l+5)},
\end{align}%
\endgroup%
where the coefficients $F_r^{[n]}$ are $l$-independent functions of $r_p$. Only the first two of these functions is needed to ensure convergence to the correct self-force, but we include additional terms to accelerate convergence of the sum.

A standard test of this procedure is that after subtracting each subsequent term $F^{[n,l]}_r$, the remainder falls off with $l$ in the characteristic manner of the next term ($F^{[n+1,l]}_r$ or $F^{[n+2,l]}_r$). We verify that our results exhibit this correct behavior in Fig.~\ref{fig:radialselfforce_lightring_ell_convergence}. Other components of the force not shown here, as well as the regular field, exhibit the same behaviour. For all these quantities, we use four orders of regularization, corresponding to the red solid line. We see the results for this curve are contaminated with numerical error at high $l$ and hence deviate from the dashed gray line (corresponding to $F^{[6,l]}_r$), but this error floor is at the level of~$10^{-7}$.

\end{document}